\documentstyle[preprint,aps,epsf]{revtex}
\begin{document}
\draft
\title{Nuclear-polarization correction to the bound-electron $g$ factor
in heavy hydrogenlike ions}

\author{A.V. Nefiodov$^{1,2}$, G. Plunien,$^1$  and G. Soff$^1$}
\address{$^1$Institut f\"ur Theoretische Physik, Technische  Universit\"at
Dresden, Mommsenstra{\ss}e 13, D-01062  Dresden, Germany  \\ 
$^2$Petersburg Nuclear Physics Institute, 188300 Gatchina,
St.~Petersburg, Russia}

\date{Received \today}
\maketitle

\widetext
\begin{abstract}
The influence of nuclear polarization on the bound-electron $g$ factor in
heavy hydrogenlike ions is investigated. Numerical calculations are
performed for the K- and L-shell electrons taking into account the dominant
virtual nuclear excitations. This determines the ultimate limit for tests of
QED utilizing measurements of the bound-electron $g$ factor in highly charged
ions.
\end{abstract}

\narrowtext
Recent high-precision experiments for measuring the bound-electron $g$ factor
in hydrogenlike carbon have reached a level of accuracy of about $2 \times
10^{-9}$ \cite{NH00,HH00}. As a consequence, this has led to a new independent
determination of the electron mass \cite{BHH2}. Via investigations of the $g$
factor of a bound electron in a highly charged ion one can probe nontrivial
effects in bound-state QED as sensitive as in high-precision Lamb shift
experiments. A further improvement in accuracy and the extension to systems
with higher nuclear charge numbers $Z$ up to hydrogenlike uranium is intended
in the near future \cite{NH00}. Studies of $g$ factors in heavy ions are of
particular importance, since they can provide a possibility for an independent
determination of the fine-structure constant \cite{SK00,GW01}, nuclear
magnetic moments \cite{GW01}, and nuclear charge radii. In order to achieve a
level of utmost precision in corresponding theoretical calculations, one has
to account for the relativistic, higher-order QED, nuclear-size,
nuclear-recoil, and nuclear-polarization corrections
\cite{BCS97,PSS97,TB00,CMY00,Sh01,Yel01,KIS01,GSh01,ShY02}. Investigations of
QED effects in heavy systems are strongly restricted by the uncertainty due to
the finite nuclear size \cite{TB00,ShY02}. In Ref.~\cite{ShGlSh}, a specific
difference has been introduced for bound-electron $g$ factors in H- and
Li-like ions, for which the uncertainty due to the nuclear-size effect can be
significantly reduced. With an apparent accuracy of $10^{-9}$ for the
bound-electron $g$ factor one could probe higher-order QED corrections even
for uranium ions, provided that nuclear polarization effects remain negligible.

In the present Letter, we evaluate nuclear-polarization corrections to the $g$
factor in hydrogenlike ions. This reduces the remaining source of
uncertainties in the prediction of nuclear effects. Moreover, we determine 
the ultimate limit of accuracy for QED tests in measurements of the
bound-electron $g$ factor in highly charged ions. 

\begin{figure}[h]
\centerline{\mbox{\epsfysize=5cm \epsffile{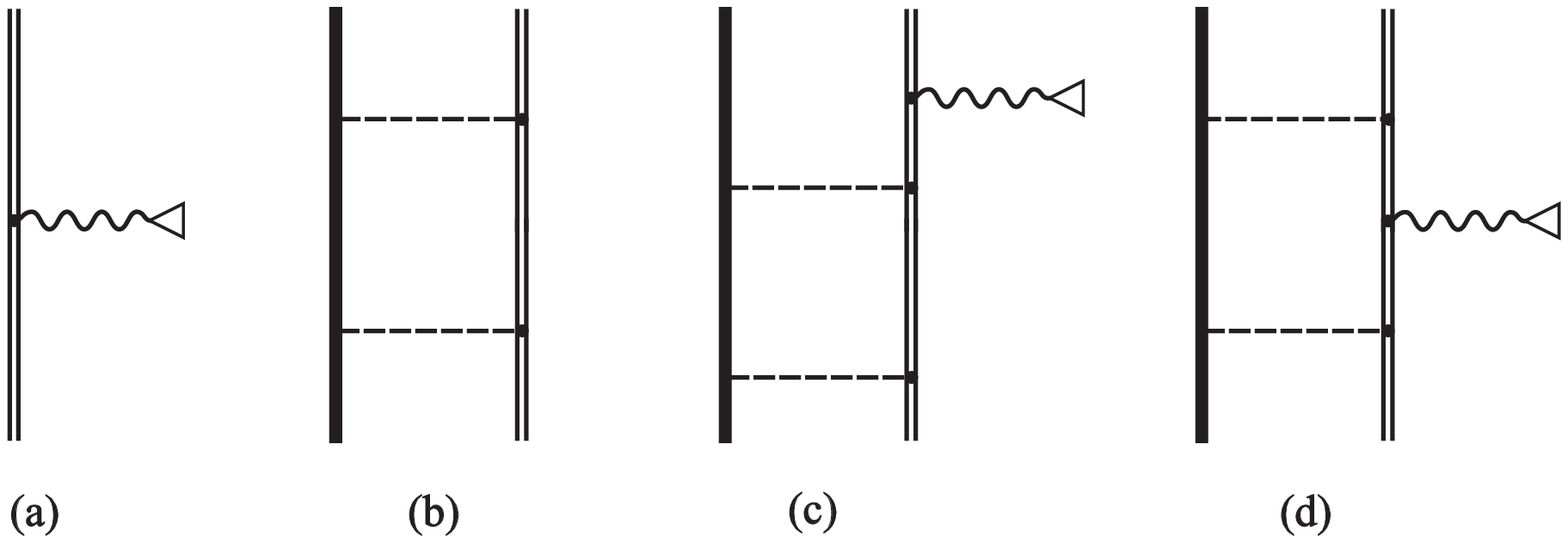}}}
\vspace*{1cm}
\caption{Diagrams representing the interaction of a bound electron 
with the external magnetic field (a), the lowest-order nuclear-polarization
effect (b), and  the nuclear-polarization correction to the bound-electron
$g$ factor (c) and (d). The heavy line denotes the nucleus. The contribution
corresponding to the graph (c) should be counted twice.}
\label{fig1}                        
\end{figure}

We consider a hydrogenlike ion with a spinless nucleus, which is placed in a
homogeneous external magnetic field $\bbox{\cal H}$ corresponding to a vector
potential $\bbox{A}(\bbox{r})=[\bbox{\cal H}\times \bbox{r}]/2$. The energy
shift of a bound-electron level $a$ within first-order perturbation theory in
the magnetic field (see Fig.~\ref{fig1}(a)) is given by ($\hbar=c=1$)
\begin{equation}
\Delta E_a = \langle \psi_a | V_{\cal H}| \psi_a \rangle  ,
\label{eq1}                                                     
\end{equation}  
where 
\begin{equation} 
V_{\cal H} = \frac{|e|}{2}(\bbox{\cal H}\cdot [\bbox{r} \times \bbox{\alpha}]) .
\label{eq2}                                                     
\end{equation} 
Choosing the $z$ axis along the direction of the field $\bbox{\cal H}$, i.e.,  
$\bbox{\cal H}=(0,0,{\cal H})$, one obtains for the energy shift
\begin{equation}
\Delta E_a = \mu_0 {\cal H} m_j g_a  ,
\label{eq3}                                                      
\end{equation} 
where $\mu_0=|e|/(2m)$ is the Bohr magneton, $m_j$ is the $z$-projection
of the total angular momentum, and $g_a$ is the bound-electron $g$ factor,
which depends on the electron configuration. In the case of a Dirac electron
in the Coulomb field of an infinitely heavy point-like nucleus, it yields
\cite{Zap79}
\begin{equation}
g^{\rm D}_{n\kappa}= \frac{\kappa}{j(j+1)} \left(\kappa \frac{\varepsilon_{n\kappa}} 
{m}   - \frac{1}{2} \right) .
\label{eq4}                                                   
\end{equation} 
Here $\kappa=(j+1/2)(-1)^{j+l+1/2}$ is the relativistic angular-momentum 
quantum number, $j$ is the total angular momentum of the electron, $l=j\pm 1/2$
defines the parity of the state, $\varepsilon_{n\kappa}$ is the one-electron
energy of the state given by
\begin{equation}
\varepsilon_{n\kappa}= m \frac{\gamma + n_r}{N}  ,
\label{eq5}                                                   
\end{equation} 
where $n_r =n - |\kappa|$ is the radial quantum number, $n$ is the principal
quantum number, $\gamma=\sqrt{\kappa^2 -(\alpha Z)^2}$, $N=\sqrt{(n_r +
\gamma)^2 +(\alpha Z)^2}$, and $\alpha=e^2$ is the fine-structure constant. Due
to various QED and nuclear effects, the observed bound-electron $g$ factor
deviates from its Dirac value (\ref{eq4}). Here we consider the 
nuclear-polarization correction $\Delta g_{n\kappa}$, which is of particular
importance in heavy ions.

The dominant contribution to the nuclear-polarization effect for heavy nuclei
arises from virtual collective nuclear excitations. Three types of collective
modes should be taken into account: (a) rotations of the deformed nuclei; (b)
harmonic surface vibrations; and (c) giant resonances. In Ref.~\cite{GPBM91},
a relativistic field theoretical approach has been developed, where  the
nuclear-polarization effects are treated perturbatively, incorporating the
many-body theory for virtual nuclear excitations within bound-state QED for
atomic electrons. The contribution of the nuclear vector current can be
omitted, because the velocities associated with collective nuclear dynamics
are nonrelativistic \cite{HaHo02}. Accordingly, one is left with the
longitudinal component of the effective photon propagator $D_{00}$ only due to
the nuclear transition  density-correlation function. In Coulomb gauge, it
can be represented in terms of a multipole decomposition as follows
\cite{GPBM91}
\widetext
\begin{equation} 
D_{00}(\bbox{r},\bbox{r}';\omega) = \sum_{L\geq 0} B(EL)\frac{2\omega_L}{2L+1}
\frac{F_L(r)F_L(r')}{\omega^2 -\omega_L^2 +i0}(\bbox{Y}_L(\Omega)\cdot 
\bbox{Y}_L(\Omega')) .
\label{eq6}                                               
\end{equation}
\narrowtext
\noindent
Here $\omega_L = E_L - E_0$ are the nuclear  excitation energies with
respect to the ground-state energy $E_0$ of the nucleus and
$B(EL)=B(EL;0\to L)$ are the  corresponding reduced electric transition
probabilities. The radial shape parametrizing the nuclear transitions is
carried by the functions
\widetext
\begin{equation}
F_L(r)=\frac{4\pi}{(2L+1)R_0^L}\left[\frac{r^L}{R_0^{L+1}}\Theta(R_0 -r) + 
\frac{R_0^L}{r^{L+1}}\Theta(r-R_0) \right]   
\label{eq7}                                                  
\end{equation}
\narrowtext
\noindent
for the case of multipole excitations with $L\geq 1$ and 
\begin{equation}
F_0(r) =\frac{5\sqrt{\pi}}{2R_0^3}\left[1 -\left( \frac{r}{R_0} \right)^2
\right] \Theta(R_0 -r)
\label{eq8}                                                  
\end{equation}
for monopole excitations, respectively. Here $R_0$ is an average radius
assigned to the nucleus in its ground state. The presence of $\Theta$-functions
in the expressions (\ref{eq7})  and  (\ref{eq8}) reflects the sharp surface
approximation for collective excitations. The form (\ref{eq6}) of
the propagator is convenient for numerical evaluations, since the parameters
characterizing the  nuclear dynamics $\omega_L$ and $B(EL)$ can be taken
from experiment. Nuclear-polarization corrections to the Lamb shift  (see
graph on Fig.~\ref{fig1}(b)) have been calculated in
Refs.~\cite{GPBM91,HaHo02,GPS95,NLP96}.

The nuclear-polarization contribution to the bound-electron $g$ factor appears
as the lowest-order nuclear-polarization correction to the diagram
\ref{fig1}(a). To first order in $V_{\cal H}$, this perturbation gives rise to
a modification of the wave function, of the binding  energy, and of the
electron propagator. The corresponding contributions are referred to as the
irreducible part, the reducible part, and the vertex  part, respectively. The
nuclear-polarization energy shift of the state under consideration may be
represented by the diagrams depicted in  Figs.~\ref{fig1}(c) and
\ref{fig1}(d). Let us consider first the energy correction due to the
irreducible part of the graph ~\ref{fig1}(c)
\widetext
\begin{eqnarray}
\Delta E_{a}^{\rm irr} &=& -2\alpha \sum_{L,M}B(EL)
\frac{2 \omega_L }{2L+1} \sum_{n,k}^{\varepsilon_k\neq\varepsilon_a}
\int \limits_{-\infty}^{+\infty}\frac{d\omega}{2\pi i} 
\frac{\langle \psi_a | F_LY_{LM} |\psi_n \rangle}{\varepsilon_a - \omega -
\varepsilon_n(1-i0)} \nonumber   \\
&\times& \frac{\langle \psi_n | F_LY^*_{LM} |\psi_k \rangle}{\omega^2  -
\omega^2_L +i0} \frac{\langle \psi_k | V_{\cal H}| \psi_a \rangle }
{\varepsilon_a  -  \varepsilon_k(1-i0)}  . 
\label{eq9}                                                
\end{eqnarray} 
\narrowtext
\noindent
Here the indices $n$ and $k$ in the sum  run over the entire Dirac spectrum. 

After integration over frequencies $\omega$ and summation over angular
projections, Eq.~(\ref{eq9}) takes the form 
\widetext
\begin{eqnarray}
\Delta E_{n\kappa}^{\rm irr} &=& \mu_0{\cal H} m_j \frac{\alpha}{2\pi}
\frac{\kappa m}{j(j+1)}  \sum_L  B(EL) \sum_{n_1, \kappa_1}
\sum_{n_2}^{n_2\neq n}  \left[C^{j_1 \frac{1}{2}}_{j \frac{1}{2} L 0} \right]^2
\nonumber  \\
&\times& \frac{\langle n\kappa |F_L|n_1\kappa_1\rangle  \langle n_1\kappa_1
|F_L |n_2\kappa \rangle} {\varepsilon_{n\kappa} - \varepsilon_{n_1\kappa_1} -
{\rm sgn}(\varepsilon_{n_1\kappa_1}) \omega_L}  \frac{\langle n_2 \kappa
|r\sigma_x | n\kappa \rangle}{\varepsilon_{n\kappa} - \varepsilon_{n_2\kappa}} .
\label{eq10}                                                   
\end{eqnarray}
\narrowtext
\noindent
The sum over $\kappa_1$ is restricted to those intermediate states,
where $l+l_1+L$ is even. In Eq.~(\ref{eq10}), a two-component radial vector
$\langle r | n\kappa \rangle$ is determined by
\begin{equation}
\langle r | n\kappa \rangle={ {P_{n\kappa}(r)} \choose {Q_{n\kappa}(r)} }  ,
\label{eq11}                                                    
\end{equation} 
where $P_{n\kappa}(r) =rg_{n\kappa}(r)$ and $Q_{n\kappa}(r) =rf_{n\kappa}(r)$, 
with $g_{n\kappa}(r)$ and $f_{n\kappa}(r)$ being the upper and lower radial
components of the Dirac wave function \cite{AB81}, respectively. The radial
matrix element is given by
\begin{equation}
\langle a |F_L |b \rangle =\int \limits_0^\infty dr  F_L(r)
\left[P_a(r)P_b(r) + Q_a(r)Q_b(r)\right]
\label{eq12}                                                  
\end{equation} 
and $\sigma_x$ denotes the Pauli matrix. The sum 
\begin{equation}
\langle r | \overline {n\kappa }\rangle =  \sum_{n'}^{n'\neq n}
\frac{\langle r | n' \kappa \rangle  \langle n' \kappa |r\sigma_x | n\kappa \rangle}
{\varepsilon_{n\kappa} - \varepsilon_{n'\kappa}}
\label{eq13}                                                   
\end{equation}  
can be evaluated analytically using the generalized virial relations for the
Dirac equation \cite{Sh91}. The upper and lower components
$\overline{P}_{n\kappa}(r)$ and $\overline{Q}_{n\kappa}(r)$ of the vector 
$\langle r | \overline {n\kappa }\rangle$ read \cite{Sh01}
\widetext
\begin{eqnarray}
\overline{P}_{n\kappa}(r)&=&\frac{1}{m^2}\left[\left(m\kappa - \frac{m}{2} + 
\kappa\varepsilon_{n\kappa}\right)r + \alpha Z \kappa\right]Q_{n\kappa}(r) +
\frac{\kappa}{2m^2}(1 -  2\kappa)P_{n\kappa}(r) , 
\label{eq14}                    \\                           
\overline{Q}_{n\kappa}(r)&=&\frac{1}{m^2}\left[\left(m\kappa + \frac{m}{2} - 
\kappa\varepsilon_{n\kappa}\right)r - \alpha Z \kappa\right]P_{n\kappa}(r) +
\frac{\kappa}{2m^2}(1 +  2\kappa)Q_{n\kappa}(r)  .
\label{eq15}                                                 
\end{eqnarray} 
\narrowtext
\noindent
Finally, the corresponding irreducible part of the correction $\Delta
g_{n\kappa}$ can be expressed as
\widetext
\begin{equation}
\Delta g^{\rm irr}_{n\kappa} = \frac{\alpha}{2\pi} \frac{\kappa m}{j(j+1)} 
\sum_L  B(EL) \sum_{n_1, \kappa_1} \left[ C^{j_1 \frac{1}{2}}_{j \frac{1}{2} L
0} \right]^2  \frac{\langle n\kappa |F_L|n_1\kappa_1\rangle \langle n_1\kappa_1
|F_L|\overline{n\kappa} \rangle} {\varepsilon_{n\kappa} -
\varepsilon_{n_1\kappa_1}  - {\rm sgn}(\varepsilon_{n_1\kappa_1}) \omega_L} .
\label{eq16}                                                 
\end{equation} 
\narrowtext

The reducible part of the graph depicted in Fig.~\ref{fig1}(c) has to be
considered together with the contributions resulting from diagrams
\ref{fig1}(a) and \ref{fig1}(b). The corresponding corrections to the energy
shift reads
\widetext
\begin{eqnarray}
\Delta E_{a}^{\rm red} &=& \alpha \langle \psi_a | V_{\cal H}| \psi_a \rangle 
\sum_{L,M} B(EL)\frac{2 \omega_L }{2L+1} \sum_{n} 
\int \limits_{-\infty}^{+\infty}\frac{d\omega}{2\pi i} 
\frac{\langle \psi_a | F_LY_{LM} |\psi_n \rangle}{\omega^2 - \omega^2_L +i0} 
\nonumber   \\
&\times& \frac{\langle \psi_n | F_LY_{LM}^{*} |\psi_a \rangle}{[\varepsilon_a -
\omega -  \varepsilon_n(1-i0)]^2} , 
\label{eq17}                                                     
\end{eqnarray}  
\narrowtext
\noindent
leading to the $g$-factor correction 
\widetext
\begin{equation}
\Delta g^{\rm red}_{n\kappa} = -\frac{\alpha}{4\pi} g^{\rm D}_{n\kappa}
\sum_L  B(EL) \sum_{n_1, \kappa_1}  \frac{\left[C^{j_1 \frac{1}{2}}_{j
\frac{1}{2} L 0}\right]^2  \langle n\kappa |F_L|n_1\kappa_1\rangle{}^2}
{[\varepsilon_{n\kappa} - \varepsilon_{n_1\kappa_1}  -{\rm
sgn}(\varepsilon_{n_1\kappa_1})\omega_L]^2} .
\label{eq18}                                                      
\end{equation} 
\narrowtext
\noindent
Here $g^{\rm D}_{n\kappa}$ is the Dirac $g$ factor given by Eq.~(\ref{eq4}). 
In Eqs.~(\ref{eq16}) and (\ref{eq18}), the sum  $l +l_1 + L$ again should be 
even. 

Let us now turn to the nuclear-polarization correction to the
vertex as depicted in Fig.~\ref{fig1}(d). The corresponding energy shift is
determined by
\widetext
\begin{eqnarray}
\Delta E_{a}^{\rm ver} &=& -\alpha \sum_{L,M} B(EL)
\frac{2 \omega_L }{2L+1} \sum_{n,k} \int
\limits_{-\infty}^{+\infty}\frac{d\omega}{2\pi i}  \frac{\langle \psi_a |
F_LY_{LM} |\psi_n \rangle}{\varepsilon_a -  \omega - \varepsilon_n(1-i0)}
\nonumber   \\ 
&\times& \frac{\langle \psi_n | V_{\cal H}| \psi_k \rangle
}{\omega^2 - \omega^2_L +i0} \frac{\langle \psi_k | F_LY^*_{LM} |\psi_a
\rangle} {\varepsilon_a - \omega - \varepsilon_k(1-i0)}  .
\label{eq19}                                                    
\end{eqnarray}  
\narrowtext
\noindent
The integration over $\omega$ and the summation over angular variables leads
to the corresponding expression for $\Delta g^{\rm ver}_{n\kappa}$, which is
conveniently represented as the sum of a pole term
\widetext
\begin{eqnarray}
\Delta g^{\rm pol}_{n\kappa}&=&  \frac{\alpha}{4\pi} \frac{\kappa
m}{\sqrt{j(j+1)(2j+1)}}  \sum_L  B(EL) \sum_{n_1, \kappa_1}
\frac{(2j_1+1)^{3/2}}{\sqrt{j_1(j_1+1)}}  \left[C^{j_1 \frac{1}{2}}_{j
\frac{1}{2} L 0}\right]^2 \left\{ {j_1 \atop j} {j_1 \atop j} {1 \atop L}
\right\} \nonumber \\ &\times & \frac{\langle n_1\kappa_1 | r\sigma_x|
n_1\kappa_1 \rangle \langle n\kappa |F_L|n_1\kappa_1\rangle{}^2
}{[\varepsilon_{n\kappa} - \varepsilon_{n_1\kappa_1}   - {\rm
sgn}(\varepsilon_{n_1\kappa_1}) \omega_L]^2}   
\label{eq20}                                                   
\end{eqnarray}
\narrowtext
\noindent
and of a residual term 
\widetext
\begin{eqnarray}
\Delta g^{\rm res}_{n\kappa}&=& \frac{\alpha}{\pi} \frac{2\kappa
m}{\sqrt{j(j+1)(2j+1)}}  \sum_{L}  B(EL) \mathop{{\sum}'}_{n_1,n_2}
\sum_{\kappa_1, \kappa_2} \sqrt{j_2 + 1/2}C^{j_1 \frac{1}{2}}_{j \frac{1}{2} L
0}C^{j_2 \frac{1}{2}}_{j \frac{1}{2} L 0} C^{j_1 \frac{1}{2}}_{j_2
-\frac{1}{2}  1 1 } \nonumber \\ 
&\times & \left\{ {j_1 \atop j} {j_2 \atop j} {1 \atop L} \right\} 
\frac{\langle n\kappa |F_L|n_1\kappa_1\rangle \langle n_2\kappa_2 |F_L|
n\kappa\rangle} {\varepsilon_{n\kappa} - \varepsilon_{n_2\kappa_2} - {\rm
sgn}(\varepsilon_{n_2\kappa_2})\omega_L} \frac{\langle n_1 \kappa_1 |r\sigma_x
| n_2\kappa_2 \rangle}{\varepsilon_{n_1\kappa_1} - \varepsilon_{n_2\kappa_2}} ,
\label{eq21}                                                
\end{eqnarray}
\narrowtext
\noindent
respectively. Here $\Delta g^{\rm pol}_{n\kappa}$ accounts for the terms 
with $n_1 = n_2$  and  $\kappa_1=\kappa_2$ in the sums over
intermediate states. The prime in the sum in Eq.~(\ref{eq21}) indicates that
$\varepsilon_{n_1\kappa_1} \neq \varepsilon_{n_2\kappa_2}$ when $\kappa_1
=\kappa_2$, i.e.,  the pole contribution is supposed to be omitted. In
Eqs.~(\ref{eq20}) and (\ref{eq21}), the value  $l+l_1 + L$ has to be even. A 
second condition in Eq.~(\ref{eq21}) is that the sum $l_1 + l_2$ should be even
as well. The total nuclear-polarization contribution to the  $g$ factor is
determined by the sum of all contributions given by Eqs.~(\ref{eq16}),
(\ref{eq18}), (\ref{eq20}), and (\ref{eq21}).

We have evaluated the nuclear-polarization correction to the $g$ factor
taking into account a finite set of dominant collective nuclear excitations
(see Table~\ref{table1}). For low-lying rotational and vibrational levels, the
corresponding nuclear parameters, $\omega_L$ and $B(EL)$, have been taken from
experiments on nuclear Coulomb excitation. In our estimates for the
contributions due to giant resonances, we utilized phenomenological
energy-weighted sum rules \cite{NLP96,RS78}. The latter are assumed to be
concentrated in single resonant states. In the present calculations,
contributions due to monopole, dipole, quadrupole, and octupole giant
resonances have been taken into account. To evaluate the infinite summations
over the entire  Dirac spectrum, the finite basis set method has been employed.
Basis functions have been generated via B splines including nuclear-size
effects \cite{JBS88}. The major  contribution to the $\Delta g_{n\kappa}$
results from the correction to the wave function (\ref{eq16}). This is due to
the fact that the matrix element of the atomic magnetic-moment operator is
saturated over the scale of atomic distances, while the influence of
nuclear-polarization is essential in the vicinity of nucleus only. In the
irreducible term, atomic and nuclear scales come into play simultaneously.

According to our numerical results, we conclude that introducing a specific
difference of $g$ factors of H- and Li-like heavy ions \cite{ShGlSh} does not
eliminate all the nuclear effects. The influence of intrinsic nuclear dynamics
becomes noticable at a level of accuracy of about $10^{-9}$ for nuclei in the
medium $Z$-range and increases up to  $10^{-6}$ in uranium. Since
nuclear-polarization effects set a natural limit up to which bound-state QED
can be tested, one is faced here with a situation similar to the one in Lamb
shift experiments. However, within the expected accuracy of $10^{-9}$ in
$g$-factor experiments with heavy highly charged ions  one may provide a tool
for probing internal nuclear structure and for testing specific nuclear models.

The authors are indebted to T.~Beier for valuable and stimulating
discussions. A.N. is grateful for financial support from RFBR (Grant No.
01-02-17246) and from the Alexander von Humboldt  Foundation. G.P. and  G.S.
acknowledge financial support from BMBF, DFG, and GSI.

\newpage
\widetext
\noindent

\newpage
\begin{table}[h]
\widetext
\noindent
\caption{Nuclear-polarization effects to the $g$ factor of K- and L-shell
electrons in hydrogenlike ions. Column (a): contributions from low-lying
rotational and vibrational nuclear modes using experimental values for nuclear
excitation energies $\omega_L$ and electric transition strengths $B(EL)$; (b)
contributions from giant resonances employing empirical sum rules [20,23];
(c) total effect. The numbers in parentheses are powers of ten.}
\label{table1}                         
\end{table}
\noindent
\begin{tabular}{llllllllllll} \hline  \hline
 & \multicolumn{3}{c}{$-\Delta g_{1s}$} & &
\multicolumn{3}{c}{$-\Delta g_{2s}$} & & 
\multicolumn{3}{c}{$-\Delta g_{2p_{1/2}}$}
\\  \cline{2-4} \cline{6-8} \cline{10-12} 
& \multicolumn{1}{c}{(a)} & \multicolumn{1}{c}{(b)} & 
\multicolumn{1}{c}{(c)} & & \multicolumn{1}{c}{(a)} & 
\multicolumn{1}{c}{(b)} & \multicolumn{1}{c}{(c)} & & 
\multicolumn{1}{c}{(a)} & \multicolumn{1}{c}{(b)} & 
\multicolumn{1}{c}{(c)}  \\ \hline
$^{\ 84}_{\ 36}$Kr & $1.0(-10)$ & $1.1(-9)$ & $1.2(-9)$ & &
$1.3(-11)$ & $1.5(-10)$ & $1.6(-10)$  & &
$1.5(-13)$ & $1.8(-12)$ & $2.0(-12)$  \\ 
$^{102}_{\ 44}$Ru & $1.2(-9)$ & $3.3(-9)$ & $4.5(-9)$ & &
$1.7(-10)$ & $4.5(-10)$ & $6.2(-10)$  & &
$3.1(-12)$ & $8.6(-12)$ & $1.2(-11)$  \\ 
$^{112}_{\ 48}$Cd & $1.4(-9)$ & $5.5(-9)$ & $6.9(-9)$ & &
$1.9(-10)$ & $7.7(-10)$ & $9.6(-10)$  & &
$4.3(-12)$ & $1.8(-11)$ & $2.2(-11)$  \\ 
$^{142}_{\ 60}$Nd & $1.7(-9)$ & $2.1(-8)$ & $2.3(-8)$ & &
$2.6(-10)$ & $3.2(-9)$ & $3.5(-9)$ &&
$9.8(-12)$ & $1.2(-10)$ & $1.3(-10)$   \\ 
$^{158}_{\ 64}$Gd & $4.7(-8)$ & $3.4(-8)$ & $8.1(-8)$ & &
$7.3(-9)$ & $5.1(-9)$ & $1.2(-8)$  &&
$3.1(-10)$ & $2.3(-10)$ & $5.4(-10)$   \\ 
$^{162}_{\ 66}$Dy & $6.0(-8)$ & $4.1(-8)$ & $1.0(-7)$ & &
$9.3(-9)$ & $6.3(-9)$ & $1.6(-8)$  &&
$4.3(-10)$ & $3.0(-10)$ & $7.3(-10)$   \\ 
$^{174}_{\ 70}$Yb & $8.6(-8)$ & $6.2(-8)$ & $1.5(-7)$ & &
$1.4(-8)$ & $9.8(-9)$ & $2.4(-8)$ &&
$7.4(-10)$ & $5.4(-10)$ & $1.3(-9)$    \\ 
$^{196}_{\ 78}$Pt & $4.5(-8)$ & $1.3(-7)$ & $1.8(-7)$ & &
$7.6(-9)$ & $2.2(-8)$ & $3.0(-8)$ & &
$5.5(-10)$ & $1.6(-9)$ & $2.2(-9)$    \\ 
$^{202}_{\ 80}$Hg & $2.1(-8)$ & $1.6(-7)$ & $1.8(-7)$ & &
$3.7(-9)$ & $2.8(-8)$ & $3.2(-8)$ & &
$2.9(-10)$ & $2.1(-9)$ & $2.4(-9)$   \\ 
$^{208}_{\ 82}$Pb & $2.2(-8)$ & $2.0(-7)$ & $2.2(-7)$ & &
$3.8(-9)$ & $3.4(-8)$ & $3.8(-8)$ & &
$3.2(-10)$ & $2.9(-9)$ & $3.2(-9)$  \\ 
$^{232}_{\ 90}$Th & $6.0(-7)$ & $4.2(-7)$ & $1.0(-6)$ & &
$1.1(-7)$ & $7.8(-8)$ & $1.9(-7)$ & &
$1.2(-8)$ & $8.4(-9)$ & $2.0(-8)$  \\ 
$^{238}_{\ 92}$U & $9.0(-7)$ & $5.0(-7)$ & $1.4(-6)$ & &
$1.8(-7)$ & $9.5(-8)$ & $2.7(-7)$  & &
$1.9(-8)$ & $1.1(-8)$ & $3.0(-8)$  \\   \hline  \hline
\end{tabular}

\end{document}